Hydrogen-substituted superconductors SmFeAsO$_{1-x}$H$_x$ misidentified as oxygen-deficient SmFeAsO$_{1-x}$

Yoshinori Muraba,[†] Soshi Iimura,[‡] Satoru Matsuishi,[†] and Hideo Hosono[*,†,‡]

*Materials Research Center for Element Strategy, Tokyo Institute of Technology, 4259 Nagatsuta-cho, Midori-ku, Yokohama 226-8503, Japan*

*Materials and Structures Laboratory, Tokyo Institute of Technology, 4259 Nagatsuta-cho, Midori-ku, Yokohama 226-8503, Japan*

*Corresponding author, Email: hosono@lucid.msl.titech.ac.jp

Phone: +81-(0)45-924-5359  Fax: +81-(0)45-924-5196

# ABSTRACT

We investigated the preferred electron dopants at the oxygen sites of 1111-type SmFeAsO by changing the atmospheres around the precursor with the composition of Sm:Fe:As:O = 1:1:1:1-$x$ in high-pressure synthesis. Under $H_2O$ and $H_2$ atmospheres, hydrogens derived from $H_2O$ or $H_2$ molecules were introduced into the oxygen sites as a hydride ion, and $SmFeAsO_{1-x}H_x$ was obtained. However, when the $H_2O$ and $H_2$ sources were removed from the synthetic process, nearly stoichiometric SmFeAsO was obtained and the maximum amount of oxygen vacancies introduced remained $x = 0.05(4)$. Density functional theory calculations indicated that substitution of hydrogen in the form of $H^-$ is more stable than the formation of an oxygen vacancy at the oxygen site of SmFeAsO. These results strongly imply that oxygen-deficient $SmFeAsO_{1-x}$ reported previously is $SmFeAsO_{1-x}H_x$ with hydride ion incorporated unintentionally during high-pressure synthesis.

**INTRODUCTION**

Iron-based superconductors have attracted much attention since their discovery in 2008, and their characteristics such as material diversity and multi-orbital nature have been clarified through intensive research.[1–6] The 1111-type compound $Ln$FeAsO ($Ln$ = lanthanide), based on alternate stacks of FeAs anti-fluorite-type conducting layers and $Ln$O fluorite-type insulating layers, is a typical parent compound for iron-based superconductors.[1,7–11] The stoichiometric 1111-type families show no superconductivity and undergo structural and magnetic transitions instead.[12–14] Superconductivity is induced when these transitions are suppressed, and the critical temperature ($T_c$) is 55 K for $Ln$ = Sm; this is the highest reported value for iron pnictides.[9,15] Although various methods for suppressing these transitions and tuning their physical properties have been proposed,[1,15–22] electron doping by hydride substitution ($O^{2-} \rightarrow H^- + e^-$)[23,24] and the introduction of oxygen vacancies ($O^{2-} \rightarrow V_o + 2e^-$),[25–27] both of which can be achieved using high-pressure synthesis, are recognized as techniques that can be used to attain the highest $T_c$ superconductivity. A phase diagram of the critical temperature vs dopant concentration is a fundamental map that connects physical properties to electronic structures.[28–30] The width of the superconducting dome observed for each $Ln$FeAsO$_{1-x}$H$_x$ is about three times wider than those for 111- or 122-type[31–33] compounds, and a double-dome structure characterized by two different parent phases at $x = 0$ and 0.5 was found for $Ln$ = La.[29] These findings suggest that precise control of dopant concentration for preparation of phase diagrams is essential for understanding the physics of 1111-type materials. Although the substituted hydrogen and the oxygen vacancy can dope electrons up to almost the same amount of carrier (~0.6 e$^-$/Fe), suppression of superconductivity has not yet been observed for oxygen-deficient $Ln$FeAsO$_{1-x}$, suggesting that their $T_c$ vs e$^-$/Fe diagrams do not entirely overlap each other.[27,34] However, the $T_c$ values of oxygen-deficient $Ln$FeAsO$_{1-x}$ agree well with those of hydrogen-substituted ones if they are plotted against their $a$-axis dimension.[24] These contradictions can be understood by assuming that unintentional hydrogen is preferentially incorporated into the oxygen vacancy (V$_O$); the contaminated hydrogen captures an electron, and then behaves as a hydride, V$_O$ + 2e$^-$ + H → H$_O^-$ + e$_{carrier}^-$, where H$_O^-$ and e$_{carrier}^-$ denote H$^-$ at the oxygen vacancy site and carrier electron, respectively.

In this study, to examine the preferred electron-dopant species in 1111-type $Ln$FeAsO and the influence of the atmosphere during synthesis on the formation of hydrogen-substituted and oxygen-deficient $Ln$FeAsO$_{1-x}$, we annealed the precursors with the composition of Sm:Fe:As:O = 1:1:1:1-$x$ under three well-controlled atmospheres, i.e., $H_2O$, $H_2$, and $H_2O$- and $H_2$-free. Under the $H_2O$ or $H_2$ atmosphere, oxygen-deficient SmFeAsO$_{1-x}$ was oxidized by $H_2$ and $H_2O$ molecules, respectively, and hydrogen-substituted SmFeAsO$_{1-x}$H$_x$ was formed. This result clearly shows that conditions without $H_2$ and $H_2O$ are required for the synthesis of oxygen-deficient SmFeAsO$_{1-x}$. However, even under an atmosphere without $H_2O$ and $H_2$, stoichiometric SmFeAsO was formed rather than oxygen-deficient SmFeAsO$_{1-x}$. We evaluated the stability of hydrogen-substituted, oxygen-deficient, and stoichiometric SmFeAsO under these three conditions using density functional theory (DFT) calculations. The results show that hydrogen-substituted SmFeAsO$_{1-x}$H$_x$ is more stable than SmFeAsO$_{1-x}$. Under $H_2O$- and $H_2$-free atmosphere, stoichiometric SmFeAsO is the most stable form. Under each condition, oxygen-deficient SmFeAsO$_{1-x}$ was less stable than the other two forms. These results strongly imply that the oxygen vacancy in the reported oxygen-deficient $Ln$FeAsO$_{1-x}$ is contaminated unintentionally with hydrogen, and hydrogen-substituted $Ln$FeAsO was misidentified as oxygen-deficient $Ln$FeAsO$_{1-x}$.

**EXPERIMENTAL**

We tried to synthesize SmFeAsO$_{1-x}$ ($x$ = 0.2, 0.4) using SmAs (laboratory synthesized), Fe$_2$O$_3$ (3N, Kojyundo Chem. Lab. Co., Ltd.), and Fe (3N, Kojyundo Chem.

Lab. Co., Ltd.), based on the equation

$$\text{SmAs} + (1-x)/3\text{Fe}_2\text{O}_3 + (1+2x)/3\text{Fe} \rightarrow \text{SmFeAsO}_{1-x}, \quad (1)$$

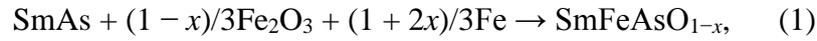

using a belt-type high-pressure anvil cell. This procedure is essentially the same as that previously reported.[25,26] A stoichiometric sample with $x = 0$ was synthesized in an evacuated silica-glass tube at 1373 K for 60 h. SmAs was prepared from its respective metals, Sm (3N, Nilaco Co., Ltd.) and As (6N, Rare Metallic Co., Ltd.) All starting materials and precursors for the synthesis were prepared in a glove box (Miwa Mfg. Co., Ltd.) filled with Ar gas ($H_2O$, $O_2 < 1$ ppm).

To achieve the aims stated in the introduction, the following three treatments were used. (i) Pre-annealing of precursors under vacuum. SmAs and $Fe_2O_3$ were annealed at 1073 K under vacuum to remove $H_2$ and $H_2O$ molecules contained and/or adsorbed on them. Table I summarizes the amounts of $H_2$ and $H_2O$ molecules. (ii) Sealing of the sample assembly for high-pressure synthesis. Figure 1(a) shows a schematic diagram of the sample cell assembly used in this study. The mixture of precursors was placed in a BN capsule, which was placed along with an $Al_2O_3$ pellet in a stainless-steel capsule with a stainless-steel cap to prevent contamination from the NaCl cell and other components, including generated gaseous species. The $Al_2O_3$ pellet provides thermal insulation and reduces heat transfer during welding from the stainless-steel cap to the sample and $H_2O/H_2$ sources, as described below. (iii) Controlling the atmosphere. An $H_2O$ or $H_2$ atmosphere was created in the welded stainless-steel tube by placing $Ca(OH)_2$ (3N, Wako Pure Chemical Industries Ltd.) or $NH_3BH_3$ (97%, Sigma-Aldrich Co., Ltd.) in it; these release $H_2O$ or $H_2$ molecules through the following decomposition reactions:

$$Ca(OH)_2 \rightarrow CaO + H_2O\uparrow, \qquad (2)$$

$$NH_3BH_3 \rightarrow BN + 3H_2\uparrow. \qquad (3)$$

The amounts of $H_2O$ and $H_2$ molecules were adjusted to twice the nominal number of oxygen vacancies $x$ in $SmFeAsO_{1-x}$. The stainless-steel capsules were heated at 1373 K and 2.5 GPa for 2 h. The crystal structures of the resulting materials were examined by powder X-ray diffraction (XRD; Bruker D8 advance TXS) using Cu Kα radiation, with the aid of Rietveld refinement using TOPAS4 code.[35] The elemental contents, except hydrogen, were determined using an electron-probe microanalyzer (EPMA; JEOL model JXA-8530F) equipped with a field-emission-type electron gun and wavelength-dispersive X-ray detectors. The micrometer-scale compositions within the main phase were probed at 10 focal points and the results were averaged, with error bars representing the two sigma standard deviation. The amounts of hydrogen incorporated into the resulting materials were evaluated using thermal desorption spectroscopy (TDS;

ESCO TDS-1000S/W). The four-probe direct current resistivity ($\rho$) was measured in the temperature range 2–200 K, using a physical properties measurement system (Quantum Design Inc.) DFT periodic calculations were performed using the generalized gradient approximation Perdew–Burke–Ernzerhof functional,[36,37] and the projected augmented plane-wave method[38] implemented in the Vienna ab initio simulation program code.[39] For stoichiometric LaFeAsO, a $\sqrt{2}a \times \sqrt{2}a \times c$ supercell was used. For oxygen-deficient and hydrogen-substituted LaFeAsO, the paramagnetic supercells summarized in Table II were used. For the gas phases of $H_2$ and $H_2O$, a molecule in a large unit cell with a lattice parameter of 1.5 nm was calculated using gamma-only calculations. The plane-basis-set cutoff energy was set at more than 800 eV. Brillouin-zone integrations were performed using a Monkhorst–Pack strategy, with k-point distances of $0.035 \text{Å}^{-1}$ or less, to calculate the total energy. The lattice parameters and atomic positions were fully relaxed by a structural optimization procedure minimizing the total energy and force.

**RESULTS**

Figure 2(a) shows the XRD patterns of $SmFeAsO_{1-x}$ with $x$ = 0.2 and 0.4 synthesized under $H_2O$ and $H_2$ atmospheres, and an atmosphere without $H_2O$ and $H_2$ molecules. The samples synthesized under these three conditions are denoted by WAT,

HYD, and NONE, respectively.

$$\text{NONE: SmAs} + (1 - x)/3\text{Fe}_2\text{O}_3 + (1 + 2x)/3\text{Fe} \rightarrow \text{SmFeAsO}_{1-x}, \quad (4)$$

$$\text{WAT: SmAs} + (1 - x)/3\text{Fe}_2\text{O}_3 + (1 + 2x)/3\text{Fe} \xrightarrow{+2x\,\text{H}_2\text{O}} \text{SmFeAsO}_{1-x}, \quad (5)$$

$$\text{HYD: SmAs} + (1 - x)/3\,\text{Fe}_2\text{O}_3 + (1 + 2x)/3\,\text{Fe} \xrightarrow{+2x\,\text{H}_2} \text{SmFeAsO}_{1-x}. \quad (6)$$

In each series, the main peaks were from the 1111 phase, and were indexed to the tetragonal ZrCuSiAs-type structure (space group $P4/nmm$). Some minor peaks were identified as reflections from SmFe$_2$AsN[40] (3.47 wt%) and SmAs (1.37 wt%) for $x = 0.4$ in the WAT series, and SmAs (1.09 wt%) for $x = 0.2$ in the HYD series. The SmFe$_2$AsN impurity was probably formed by reaction between the precursors and the BN capsule, which are in direct contact with each other during the high-pressure synthesis. For the NONE series, the amount of SmFeAsO phase was smaller than those in the WAT and HYD series, and impurity phases of SmAs and Fe gradually segregated with increasing $x$. The nominal $x$ dependences of the weight percentages of the 1111 phase, estimated by Rietveld analyses, are plotted for each series in Fig. 2(b). In the case of the NONE series only, the amount of 1111 phase decreased to ~60 wt% at $x = 0.4$. Figure 2(c) shows the variations in the lattice parameters with $x$. In both the WAT and HYD series, the $a$- and $c$-axis lengths decreased almost linearly as $x$ increased, whereas in the NONE series, they remained almost constant with changes in $x$ (0.2 and 0.8 pm decreases for $a$ and $c$, respectively, at $x = 0.4$). These results are consistent with the formation of substituted phases in the first two cases and non-formation in the third case.

Figure 3(a)–(c) show the temperature dependences of the electrical resistivity. An anomaly at ~140 K ($T_{anom.}$) in the $\rho$–$T$ curves caused by a structural transition was completely suppressed and a sudden drop in the resistivity as a result of superconductivity was observed at around 50–55 K for the WAT and HYD series. In contrast, for the NONE series, the anomaly remained, even at $x = 0.4$, and its temperatures, 138 and 133 K at $x = 0.2$ and 0.4, respectively, are comparable to that of stoichiometric SmFeAsO synthesized under ambient pressure ($T_{anom.}$ ~139 K). No superconducting transition was observed for any value of $x$ above 2 K.

The $x$ dependences of the oxygen ([O]) and hydrogen ([H]) concentrations normalized by the molar content of Sm are shown in Fig. 4(a)–(c), along with their sums ([O] + [H]). For the WAT and HYD series, the amount of oxygen decreases linearly with increasing nominal $x$. For the WAT series, the determined oxygen content is larger than the nominal content (gray line), indicating that oxygen was incorporated from the $H_2O$ atmosphere. For the HYD series, the determined oxygen content is almost the same as the nominal oxygen content. However, for both series, the resulting samples contain large amounts of hydrogen, and the hydrogen contents increase with increasing nominal $x$. The sum of the oxygen and hydrogen contents remains close to unity over the entire nominal $x$ region, indicating that hydrogen-substituted SmFeAsO$_{1-x}$H$_x$ was formed under the H$_2$O and H$_2$ atmospheres without the formation of significant amounts of oxygen vacancies. For the WAT series, both hydrogen and oxygen occupy the oxygen vacancy site in SmFeAsO$_{1-x}$ via H$_2$O incorporation from the atmosphere. For the HYD series, hydrogen occupies only oxygen vacancy site, without removing oxygen, in SmFeAsO$_{1-x}$, because the determined and nominal oxygen contents agree well. In contrast, for the NONE series, the amount of oxygen hardly changed, from 0.98(4) at $x = 0$ to 0.95(4) at $x = 0.2$ and 0.4. The amount of hydrogen detected is therefore negligibly small (less than 0.002) for each $x$.

**DISCUSSION**

First, we discuss the compounds that were formed in the WAT and HYD series. The compositional analyses suggest the formation of hydrogen-substituted SmFeAsO$_{1-x}$H$_x$. To verify this assumption, we compared the lattice parameters and $T_c$ values for both series with those for the reported hydrogen-substituted SmFeAsO$_{1-x}$H$_x$. Figure 5(a)–(c) show the lattice parameters and $T_c$ values as a function of the hydrogen content determined by TDS in the WAT (blue triangles) and HYD (red squares) series along with those for the reported hydrogen-substituted SmFeAsO$_{1-x}$H$_x$ (gray lines and circles). It is evident that all the data for both series fall well on a master curve for the hydrogen-substituted SmFeAsO$_{1-x}$H$_x$. This agreement reinforces the assumption that hydrogen-substituted SmFeAsO$_{1-x}$H$_x$ was formed by incorporation of hydrogen and oxygen supplied from H$_2$O and/or H$_2$ atmospheres for the WAT and HYD series. Moreover, these results clearly indicate that H$_2$- and H$_2$O-free conditions are required for the synthesis of oxygen-deficient SmFeAsO$_{1-x}$. What was formed in the NONE series? Can oxygen vacancies be introduced into SmFeAsO? The oxygen content determined using EPMA indicates that only the content of introduced vacancies appears to be within the error bars in the experiment, and the lattice parameters and $T_{anom}$ remain almost unchanged with increasing nominal $x$. It is therefore reasonable to conclude that nearly stoichiometric SmFeAsO was formed under an atmosphere without H$_2$O and H$_2$ molecules. These experimental results can be summarized in the following reactions:

NONE: SmAs + (1 − $x$)/3Fe$_2$O$_3$ + (1 + 2$x$)/3Fe → (1 − $x$)SmFeAsO + $x$SmAs + $x$Fe,  (7)

WAT: SmAs + (1 − $x$)/3Fe$_2$O$_3$ + (1 + 2$x$)/3Fe + $x$/2 H$_2$O → SmFeAsO$_{1-x/2}$H$_{x/2}$,  (8)

HYD: SmAs + (1 − $x$)/3Fe$_2$O$_3$ + (1 + 2$x$)/3Fe + $x$/2 H$_2$ → SmFeAsO$_{1-x}$H$_x$.  (9)

Second, we examined the validity of the above conclusion in terms of thermodynamic instability in the oxygen-deficient SmFeAsO$_{1-x}$ using DFT calculations.

This instability is evaluated as the energy differences of reaction formulas (10), (11), and (12) as follows;

$$SmFeAsO_{1-x} \rightarrow (1-x)SmFeAsO + xSmAs + xFe, \quad (10)$$

$$SmFeAsO_{1-x} + x/2\, H_2O \rightarrow SmFeAsO_{1-x/2}H_{x/2} + x/4\, H_2, \quad (11)$$

$$SmFeAsO_{1-x} + x/2\, H_2 \rightarrow SmFeAsO_{1-x}H_x, \quad (12)$$

which are calculated by the subtraction from reaction formulas (7), (8), and (9) to (1), respectively. Here, we use La instead of Sm, to decrease the calculation costs stemming from spin polarization and evaluation of the effective on-site Coulombic interaction caused by localized and unoccupied 4f electrons in Sm. Figure 6 shows the results of the energetic subtractions, from right to left, for reaction reactions (10), (11), and (12). In each case, the energy differences are negative and the magnitude increases with increasing $x$.[41] These results therefore confirm that oxygen-deficient $SmFeAsO_{1-x}$ decomposes to stoichiometric SmFeAsO for any $x$, and under realistic conditions with a finite amount of $H_2O$ or $H_2$, the formation of $SmFeAsO_{1-x}H_x$ is the most exothermic reaction. These calculated results are compatible with the experimental results.

Third, to examine the effect of the introduction of vacancy or hydrogen in oxygen-site on their electronic structures, we calculated the total DOS, PDOS and the partial electron density of $LaFeAsO_{0.75}$ and $LaFeAsO_{0.75}H_{0.25}$. Figure 7 (a) shows the paramagnetic $\sqrt{2}a \times \sqrt{2}a \times c$ supercell used for DFT calculations, in which an oxygen at (1/2, 1/2, 1/2) is replaced by hydrogen or removed. Figure 7 (b) shows partial electron density ($E$ = -6.0 to 0 eV) projected on the (1 1 0) plane of LaFeAsO$_{0.75}$ and LaFeAsO$_{0.75}$H$_{0.25}$. Obviously, the electron density at the V$_O$ is quite low in comparison with that around oxygen or hydrogen. Also, we confirmed that electron density (E < -6.0) is quite low. These results indicate that the surplus electron originating from the introduction of oxygen vacancy is doped in Fe bands located around $E_F$, i. e., O$^{2-}$ → V$_O$ + 2$e^-$/Fe. On the other hand, in the case of LaFeAsO$_{0.75}$H$_{0.25}$, the hydrogen 1$s$ level forms a deep band located at -6 to -3 eV ( close to oxygen 2$p$ location) and captures 1.74 $e^-$/H, that is, O$^{2-}$ → H$^-$ + $e^-$/Fe. Therefore, for the introduction of the oxygen vacancy, two electrons are added to the $E_F$, while for the hydrogen substitution, one electron is put in the $E_F$ and the other is trapped by the hydrogen. As a consequence the electronic energy gain is obtained when the hydrogen is incorporated into the oxygen vacancy site.

Next, we discuss where the H$_2$O and H$_2$ come from in the conventional synthesis of oxygen-deficient $Ln$FeAsO$_{1-x}$. It is natural to suppose from the present experiment that the sources were derived from precursors and/or the high-pressure cell assembly. We found by TDS measurements (see Table I) that SmAs and Fe$_2$O$_3$, which were used as precursors to synthesize SmFeAsO$_{1-x}$ in this and earlier studies, can be sources of H$_2$ and H$_2$O. In the case of SmAs, the Sm metal contained 2–7 × 10$^{-5}$ mol g$^{-1}$ H$_2$ molecules. The amount of H$_2$O in Fe$_2$O$_3$ strongly depended on the supplier and sample lot, and varied from 1 × 10$^{-6}$ to 1 × 10$^{-4}$ mol g$^{-1}$. In addition to these precursors, a pyrophyllite, Al$_2$Si$_4$O$_{10}$(OH)$_2$, which is used as a pressure medium, could also be the H$_2$O source, because dehydration occurs at ~773 K (see Fig. 1). This effect is particularly significant for cubic anvil-type high-pressure apparatus, because pyrophyllite is located just outside the graphite heater and is heated nearly to the synthesis temperature.[42] If oxygen-deficient $Ln$FeAsO$_{1-x}$ is synthesized under conventional high-pressure synthetic conditions, it is therefore unsurprising that $Ln$FeAsO$_{1-x}$H$_x$ unintentionally substituted with hydrogen is obtained.

Finally, we would like to point out the wide applicability of hydrogen substitution into oxygen sites for electron doping of oxides. Electron doping into transition-metal oxides induces various electromagnetic functionalities. Electron doping has been almost exclusively performed by aliovalent substitution at cationic sites. We reported successful heavy electron doping into LaMnAsO[43] and Sr$_2$VO$_4$,[44,45] which was unsuccessful using the conventional method, by applying this method. Heavy electron doping is important in finding new superconductors and in elucidation of pairing mechanisms. We expect that this method will be helpful in revisiting candidate materials for which $T_c$ induction has not yet been successful.[46]

**CONCLUSIONS**

So far, two effective methods have been reported for doping carrier electrons into 1111-type $Ln$FeAsO, i.e., hydrogen anion ($H^-$) substitution and introduction of oxygen vacancies at oxygen sites. However, we noted that the $T_c$ vs the lattice dimension relationships for both methods are very similar, suggesting the possibility that oxygen vacancy site in $Ln$FeAsO$_{1-x}$ samples are primarily occupied by hydrogen anions. To clarify the situation, we experimentally examined the preferred electron-dopant species at oxygen site in 1111-type $Ln$FeAsO by changing the atmospheres (H$_2$, H$_2$O, and H$_2$- and H$_2$O-free) around the precursor with the composition of Sm:Fe:As:O = 1:1:1:1-$x$ under high pressure. Special attention was paid to cell assembly to prevent external contamination. The SmFeAsO system was chosen because it has the highest $T_c$ among bulk iron-based superconductors. The results are summarized as follows.

(1) The samples synthesized under an H$_2$ or H$_2$O atmosphere in high-pressure synthesis were SmFeAsO$_{1-x}$H$_x$, although the starting materials were adjusted to form SmFeAsO$_{1-x}$.

(2) The samples with the nominal composition SmFeAsO$_{1-x}$ synthesized under H$_2$- and H$_2$O-free atmospheres were nearly stoichiometric SmFeAsO. The maximum $x$ in the resulting products remained < 0.05.

(3) DFT calculations showed that the hydrogen-substituted samples are more stable than the oxygen-vacancy-substituted samples.

(4) It is most likely that the samples of $Ln$FeAsO$_{1-x}$ reported so far are actually $Ln$FeAsO$_{1-x}$H$_x$, formed by incorporating hydrogen from the atmosphere and/or starting materials.

(5) We suggest that electron doping by hydrogen substitution at oxygen site could be widely applied to oxides and oxide-layer-bearing systems if experimental condition is properly set.

ACKNOWLEDGMENTS

We thank Emer. Prof. O. Fukunaga, Assoc. Prof. T. Atou, Assoc. Prof. T. Tada, Special Assoc. Prof. H. Mizoguchi, Dr. T. Hanna, and Dr. J. Bang for valuable discussions. This study was supported by the MEXT Element Strategy Initiative Project to form a research core.

FIGURES and TABLES caption

Figure 1. Schematic diagrams of sample cell assemblies for (a) belt-type high-pressure apparatus used in this study, and (b) cubic-type high-pressure apparatus used in conventional synthesis of *Ln*FeAsO.[25–27]

Figure 2. (a) Powder XRD patterns, (b) weight fractions, and (c) lattice constants $a$ (blue squares) and $c$ (red circles) as function of nominal $x$ of SmFeAsO$_{1-x}$ synthesized under H$_2$O (WAT) and H$_2$ (HYD) atmospheres, and atmosphere without H$_2$O and H$_2$ (NONE).

Figure 3. Temperature dependences of electrical resistivities of oxygen-deficient SmFeAsO$_{1-x}$ samples synthesized under (a) H$_2$O atmosphere (WAT), (b) H$_2$ (HYD) atmosphere, and (c) atmosphere without H$_2$O and H$_2$ (NONE).

Figure 4. Chemical composition data obtained by EPMA and TDS for SmFeAsO$_{1-x}$ synthesized under (a) H$_2$O atmosphere (WAT), (b) H$_2$ (HYD) atmosphere, and (c) atmosphere without H$_2$O and H$_2$ (NONE). The contents of oxygen [O], hydrogen [H], and their sum [O + H] are represented by blue diamonds, red triangles, and green circles, respectively. Grey lines and black dashed lines show nominal oxygen ($x$) and nominal oxygen vacancy contents ($1 - x$), respectively.

Figure 5. Lattice parameters ($a$ and $c$) and critical temperature ($T_c$) of SmFeAsO$_{1-x}$ synthesized under H$_2$O (blue triangles) and H$_2$ (red squares) atmospheres as a function of $x$ determined by TDS, superimposed with values reported for SmFeAsO$_{1-x}$H$_x$ (gray lines and circles, Ref. 10)

Figure 6. Calculated energy differences corresponding to chemical reactions (10) (green diamonds), (11) (red circles), and (12) (blue triangles).

Figure 7 DFT calculations on model structures. (a) Assumed structure for LaFeAsO$_{0.75}$ and LaFeAsO$_{0.75}$H$_{0.25}$ (b) Partial electron density ($E$ = -6.0 to 0 eV) projected on the (1 1 0) plane and (c) Total DOS, atomic PDOS.

Table 1. H$_2$ and H$_2$O contents of starting materials and sample assemblies before and after vacuum annealing at 1073 K. LOD = limit of detection. Contents were estimated using TDS.

Table 2. Supercell size used for DFT calculations of LaFeAsO$_{1-x}$ and LaFeAsO$_{1-x}$H$_x$

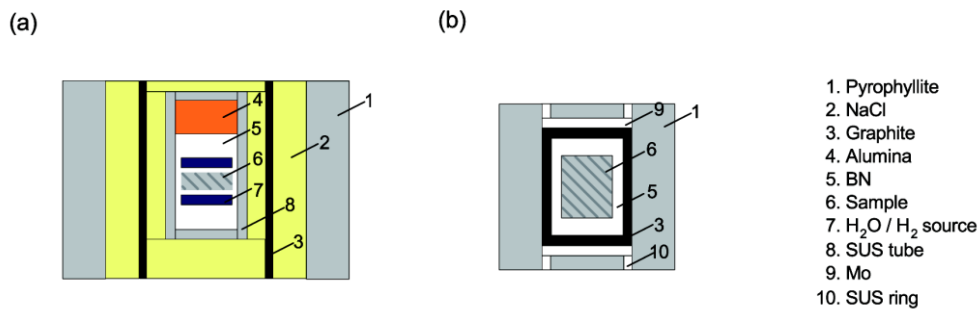

1. Pyrophyllite
2. NaCl
3. Graphite
4. Alumina
5. BN
6. Sample
7. $H_2O$ / $H_2$ source
8. SUS tube
9. Mo
10. SUS ring

Fig. 1

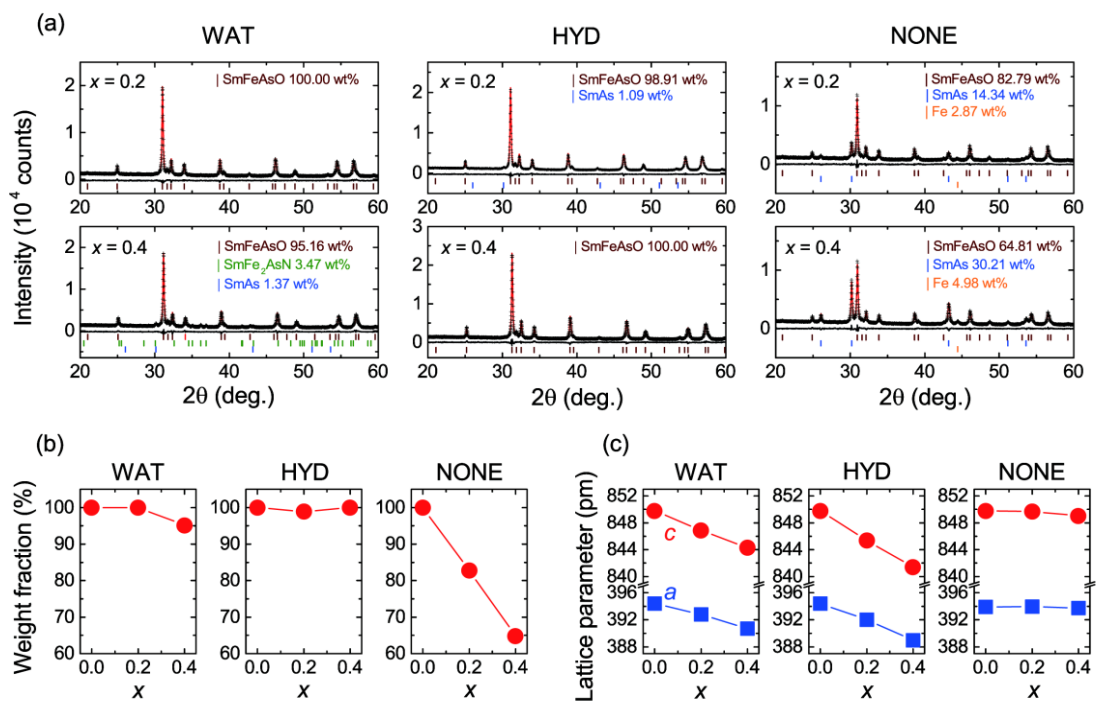

Fig. 2

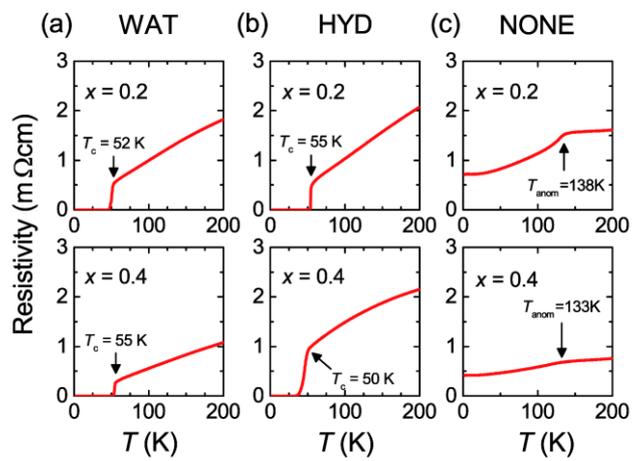

Fig. 3

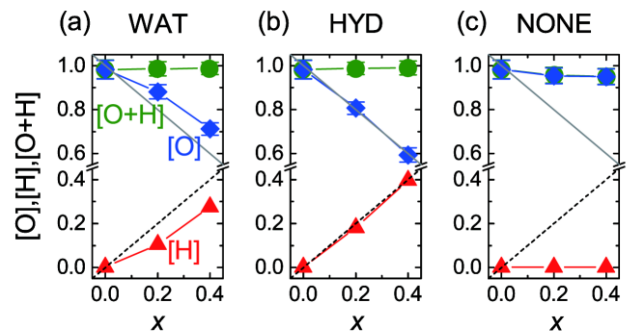

Fig. 4

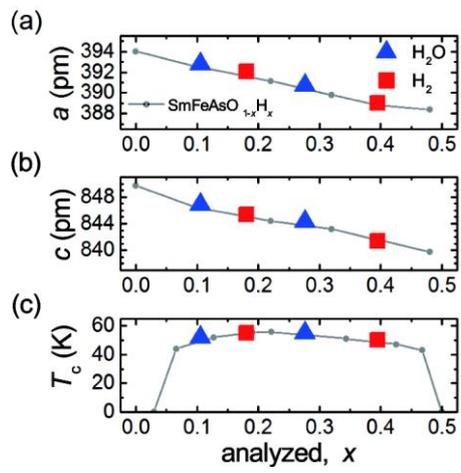

Fig. 5

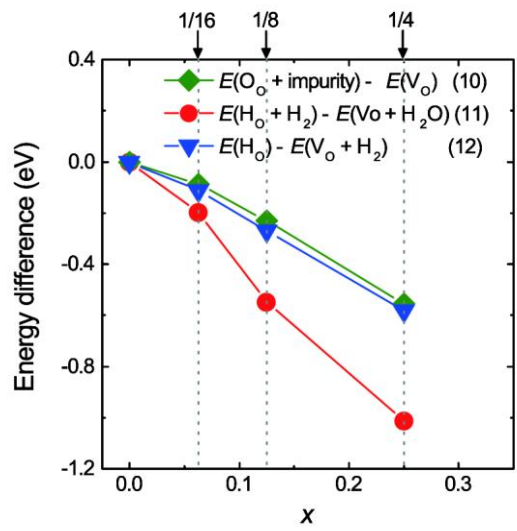

Fig. 6

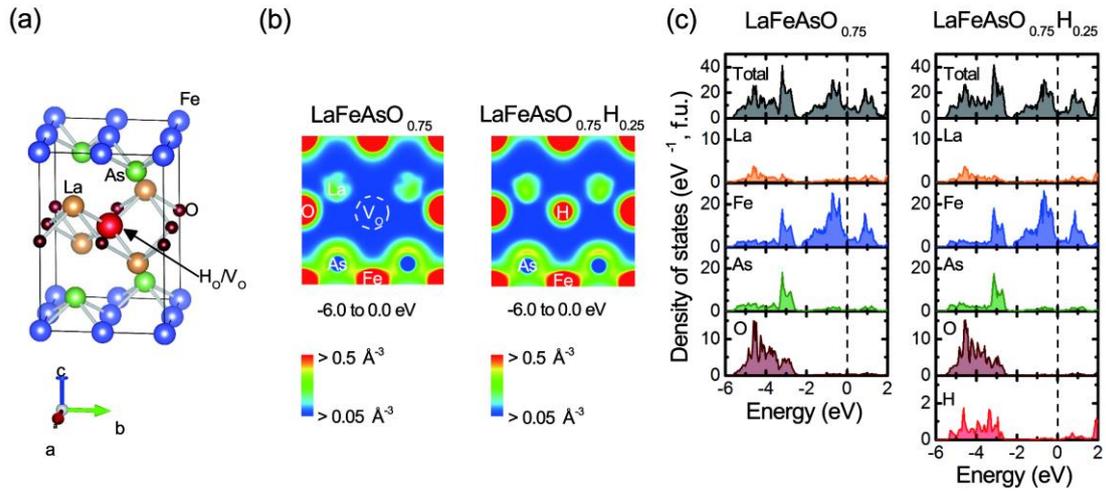

Fig. 7

Table 1

|  | Before | | After | |
| --- | --- | --- | --- | --- |
|  | $H_2$ (mol/g) | $H_2O$ (mol/g) | $H_2$ (mol/g) | $H_2O$ (mol/g) |
| SmAs | $1.78 \times 10^{-5}$ | LOD | $0.888 \times 10^{-5}$ | LOD |
| $Fe_2O_3$ | $85.8 \times 10^{-5}$ | LOD | LOD | LOD |
| Fe | $0.720 \times 10^{-5}$ | LOD | LOD | LOD |
| BN | LOD* | LOD | - | - |

Table 2

| Compound | Supercell |
|---|---|
| LaFeAsO$_{3/4}$ | $\sqrt{2}a \times \sqrt{2}a \times c$ |
| LaFeAsO$_{3/4}$H$_{1/4}$ | $\sqrt{2}a \times \sqrt{2}a \times c$ |
| LaFeAsO$_{7/8}$ | $\sqrt{2}a \times \sqrt{2}a \times 2c$ |
| LaFeAsO$_{7/8}$H$_{1/8}$ | $\sqrt{2}a \times \sqrt{2}a \times 2c$ |
| LaFeAsO$_{15/16}$ | $2a \times 2a \times 2c$ |
| LaFeAsO$_{15/16}$H$_{1/16}$ | $2a \times 2a \times 2c$ |
| LaFeAsO$_{31/32}$H$_{1/32}$ | $2\sqrt{2}a \times 2\sqrt{2}a \times 2c$ |